\documentstyle[11pt]{article}
\newcommand{\be}{\begin{equation}}
\newcommand{\ee}{\end{equation}}
\newcommand{\ba}{\begin{eqnarray}}
\newcommand{\ea}{\end{eqnarray}}

\begin{document}

\title{\vspace{-3.0cm} \hspace{0.0cm} \hspace*{\fill} \\[-5.5ex]
\hspace*{\fill} {\normalsize September 28,1995} \\
\hspace*{\fill} {\normalsize LAUR-95-3232} \\[1.5ex]
{\huge {\bf Hydrodynamical analysis of
single inclusive spectra and Bose-Einstein correlations
for $Pb+Pb$ at 160 AGeV}}}
\author{B.R. Schlei$^1$\thanks{%
E. Mail: schlei@t2.LANL.gov} , U. Ornik$^2$\thanks{%
E. Mail: ornik@tpri6o.GSI.DE} , M. Pl\"umer$^3$\thanks{%
E. Mail: pluemer@Mailer.Uni-Marburg.DE} , D. Strottman$^1$\thanks{%
E. Mail: dds@LANL.gov}{\ } and R.M. Weiner$^3$\thanks{%
E. Mail: weiner@Mailer.Uni-Marburg.DE}}
\date{$^1$Theoretical Division, Los Alamos National Laboratory, Los Alamos, NM
87545\\ $^2$GSI Darmstadt, Darmstadt, Germany\\ $^3$Physics Department,
Univ. of Marburg, Marburg, Germany }
\maketitle

\begin{abstract}
We present the first analysis
of preliminary data for  $Pb+Pb$ at 160 $AGeV$
using 3+1-dimensional relativistic hydrodynamics.
We find excellent agreement with the rapidity spectra of negative
hadrons and the correlation measurements. The data indicates a
large amount of stopping;
$65\%$ of the invariant energy of the collision is thermalized and
$73\%$ of the baryons are contained in the central fireball.  Within
our model this implies that a quark-gluon-plasma of lifetime 3.4 $fm/c$
was formed.
\end{abstract}

\hbadness=10000 \setcounter{page}{1} 

\vspace{-0.5cm}

\newpage

In the ongoing quest for a quark-gluon plasma, experiments are being pursued
using ever higher energies or masses in order to increase the lifetime of
the system either by increasing the initial energy density or the size of
the system. The probability of preparing a strongly interacting system that
shows thermodynamical behaviour and therefore is treatable by well known
thermodynamical or fluid dynamical methods increases with the size of the
system. Recently, preliminary results were announced from the NA49
Collaboration \cite {NA49_1,NA49_2,QM95} for $Pb + Pb$ central collisions at
160 $AGeV$.
The application of Bjorken's estimate \cite{bjorken} of the initial
energy density yields a value of $3\ GeV/fm^3$\cite{NA49_1}. Of course,
this estimate which is based on the value of one single observable
(the rapidity density of the transverse energy in the central region)
is no substitute for a comprehensive hydrodynamical study that takes into
account all available data on spectra and correlation functions for
different particle species.
In this paper we present results of the first analysis of $Pb+Pb$ at
160 $AGeV$ using relativistic fluid dynamics and assuming an equation-of-state
containing a  phase transition.

A fully three-dimensional solution of the hydrodynamical relativistic
Euler-equations was obtained with the computer code HYLANDER \cite{udo}. With
this solution it is possible to  reproduce \cite{jan} simultaneously mesonic
and baryonic rapidity and transverse momentum spectra of the $S+S$ reaction
at 200 $AGeV$ \cite{wenig}. Corresponding measurements had been performed
by the NA35 Collaboration. Based on these fits to the measured spectra,
predictions were made for Bose-Einstein correlation
functions\cite{bernd2,bernd3}. Those predictions agree  quantitatively with
the measurements\cite{QM95,alber}. The model also reproduces the photon data
for $S+Au$ collisions at SPS energies\cite{axel} and gives a simple
explanation for the ``soft-$p_{\perp }$ puzzle'' \cite{udo91} and the complex
behaviour of the radii extracted from pion and kaon correlations and
explained the difference in the extracted radii for pions and kaons in terms
of a cloud of pions stemming from resonance decays, which surrounds the
fireball (pion halo) \cite{bernd2}. In this paper we pursue a similar
description for
$Pb+Pb$ at 160 $AGeV$ reaction.

HYLANDER uses an initial scenario located between the extremes of
the Landau\cite{landau} and the Bjorken\cite{bjorken} initial conditions. One
has to specify an equation of state and a set of parameters which describe
the initial conditions. Our equation of state exhibits a first order phase
transition at a critical temperature $T_C=200\:MeV$ ({\it cf.}
ref.\cite{redlich}).

In the following we briefly sketch the basic features of the model which
was introduced in ref. \cite{jan}. We assume that an initial reaction of two
baryonic fluids leads to a deceleration of the two baryonic currents and to
the spread of their width in momentum space. The kinetic energy is converted
into internal excitation (thermal energy) of a third fluid
which is created in the central region. The parameters determine the
position of the maximum, the width in momentum space of the two baryonic
fluids after the collision and the spatial extent of the
central fireball. All other quantities are obtained from energy-momentum
conservation (see table 1). The parameters are determined by fitting
the pion rapidity spectrum; all other spectra are predictions of the model.

In Fig. 1a is shown the measured rapidity spectrum from the NA49 Collaboration
\cite{NA49_1,NA49_2} of negative hadrons for $Pb + Pb$ central collisions at
160 $AGeV$. In table 1 we show the parameter set and the related physical
quantities of the initial fireball as described in ref. \cite{jan} and
compare them to the $S+S$ case. The $Pb+Pb$ system leads to a larger
stopping than does $S+S$ and this results also in the increase
of the inelasticity in $Pb+Pb$ compared to $S+S$.

In the following we discuss some of the results for the single- and
double-inclusive spectra of hadrons and mesons. All calculations are based
on thermal as well as on chemical equilibrium. In both types of spectra we
include the effect of resonance decays. The influence of partial coherence
\cite{bernd3} will not be considered here. A detailed analysis $Pb+Pb$ at
160 $AGeV$ will be published in ref.\cite {future}.

Consistent with expectation the degree of stopping and thermalization is
higher in $Pb+Pb$ and the amount of thermal energy in the central fireball
increases from 43\% $(S+S)$ to 65\% $(Pb+Pb)$. Due to the higher stopping,
73\% of the baryons are located now in the central region compared to only
49\% in the $S+S$ case. The location of the maximum density of the two
baryon currents $y_m$ in rapidity space is significantly shifted into the
central rapidity region. It is important to note that the baryons for $
Pb+Pb$ are almost stopped. The resulting high baryonic density of 2.14
$fm^{-3}$ in the center and 3.2 $fm^{-3}$ in the forward region is three
times higher than in $S+S$ and can therefore no longer be neglected. It will
have a strong influence on strangeness, photon and lepton production as well
as on the relative meson abundancies.

The maximum lifetime of the fireball is increased from 6.9 $fm/c$ $(S+S)$ up
to 14.5 $fm/c$ for $Pb+Pb$.
We observe approximately the same behaviour for the lifetimes
of the QGP.  Whereas in $S+S$ the lifetime of the QGP was short (1.5 $fm/c$),
in the $Pb+Pb$ case  a QGP persists for 3.4 $fm/c$ \footnote{ We mention that
even for those who do not believe in hydrodynamics, this value is an upper
limit.}. Taking into account that the initial volume of the $Pb+Pb$ system is
$\sim$ eight times larger than in the $S+S$ case, we note that the lifetime
increases slower than does the volume. This effect is due to the cooling by
transverse rarefaction waves which are only sensitive to the difference in
the transverse radius which differs approximately by a factor two from
$S+S$ to $Pb+Pb$.

For $Pb+Pb$ at 160 $AGeV$ we find average values for the baryonic
chemical potential $\langle \mu _B\rangle =363\:MeV$ and for the strangeness
chemical potential $\langle \mu _S\rangle =69\:MeV$ using $T_f=139\:MeV$ for
the freeze-out temperature. As a result we obtain a chemical composition of
$48.9\%$ thermal (directly emitted) pions, $15.4\%$ thermal kaons, $ 21.6\% $
mesons (such as $\rho $, $\omega $, $\eta $, $K^{\star }$, ... etc.), $13.7\%$
baryons (such as $p$, $\Delta $, $\Sigma$, $\Lambda $, ... etc.) and $0.4\%$
anti-baryons (such as $\bar p,\bar \Delta$, ... etc.).
The detailed sub-fractions will be given in ref.\cite{future}.
The corresponding values for $S+S$ at 200 $AGeV$ are given in ref.\cite{jan}.

The calculation of single particle inclusive spectra with HYLANDER was
extensively discussed in ref.\cite{jan}. Fig. 1a shows our model predictions
for
negative thermal $\pi^-$ compared to all negative $\pi$.
We  stress that due to our choice of the freeze-out temperature of
$T_f=139\:MeV$,
there are large resonance contributions ($40\%-50\%$) to the pionic spectra.
The rest of our rapidity spectrum of negative hadrons $h^-$
is made up by contributions from negatively charged kaons. The $K^-$ spectrum
consists mainly of directly emitted kaons; less than $10\%$ come from the $
K^\star$ resonance.

The $h^-$ rapidity spectrum has been chosen to fit the NA49 data in such a
way that the proton rapidity spectrum has the shape observed in
the NA44 experiment\cite{jacak}. Since it is not clear from the NA44
experiment to what extent $\Lambda$ decays appear in the proton production,
we show our results for the proton rapidity spectra with and without
contributions from the $\Lambda$ decay. Within these limits we reproduce the
results for the NA44 proton rapidity spectra.

Transverse momentum spectra are shown in Fig. 1b. They were not involved in
finding a parameter set for the fit of experimental data and are therefore
predictions.

The calculation of Bose-Einstein correlations (BEC) was performed using the
formalism outlined in refs. \cite{bernd2,bernd3,bernd1} including the decay
of resonances. The hadron source is assumed to be fully chaotic.
We present results for both pion and kaon BEC.

In the previous section we mentioned that in the case of pion
interferometry we have to deal with a large fraction -- $40\%$ to $50\%$ -- of
the pions originating from resonances. The effects of resonance decays on
two-particle Bose-Einstein correlation functions are shown in Fig. 2. We
show examples of correlation functions in the longitudinal and transverse
directions.
The $K$ and $q$ variables indicate our choice of $K^\mu=\textstyle%
{\frac{1}{2}}(k_1^\mu+k_2^\mu)$ and $q^\mu=k_1^\mu-k_2^\mu$ to be the
average and the relative 4-momentum of the particle pair with 4-momenta $%
k_1^\mu$ and $k_2^\mu$. The contributions from resonance decays are
successively added to the correlation function of thermal $\pi^-$.

By adding several interferometry contributions from resonance decays, the
correlation functions become narrower reflecting a larger effective source
size. We point out that the effect of the $\omega$ resonance is clearly
visible within experimental resolution ($\sim 5 \:MeV/c$),  although its
lifetime $\tau _\omega $ is rather large ($\tau _\omega =23.4\:fm/c$)
compared to the spatial dimensions of the fireball\footnote{This is in
contrast  to ref.\cite{csoergo} where the contribution of the $\omega$
resonance was found to be invisible.}.  The decay of the $\omega$ resonance
is also mainly responsible for the deviation of the functional shape of the
correlation function from an approximately Gaussian shape.  Another
important feature of resonance decay effects is the appearance of $\eta $
decays.  Due to the very large lifetime ($\tau _\eta =1.64\times 10^5\:fm/c$)
of the $\eta$ resonance, the two-particle correlation function becomes so
narrow that the result is a decrease in the experimental resolvable
intercept to values below 1.80, depending on the experimental
resolution\footnote{ The true intercept is still two for the single
correlation function, i.e. a correlation function which contains thermal
as well as resonance contributions.} ({\it cf.} \cite{bernd3}).

In order to extract effective hadron source radii, we fit our results to the
Gaussian form used by experimentalists (for the choice of the variables, cf.,
e.g. ref.\cite{bernd3}):
\begin{equation}
\label{eq:fit}C_2(\vec{k}_1,\vec{k}_2)\:=\: 1\:+\:\lambda\:\exp
\left[-\frac{ 1}{2} (q_\parallel^2 R_\parallel^2 \:+\:q_{side}^2
R_{side}^2\:+\:q_{out}^2 R_{out}^2) \right] \:.
\end{equation}
It should be emphasized that here $\lambda$ does {\it not} represent the
effect of coherence, but the effective reduction of the intercept due to the
contributions from $\eta$ decays.

Fig. 3a shows our predictions for effective radii $R_\parallel$, $R_{side}$
and $R_{out}$ as functions of rapidity and transverse momentum of the pair,
both for $\pi^-\pi^-$ (solid lines) and for $K^-K^-$ pairs (dashed lines).
For comparison, we have also included the curves for thermally produced
pions (dotted lines). In the case of interferometry of negative kaons, we
checked that the $K^\star $ resonance gives negligible contributions to
the two-particle BECs. It is also seen that the direct (thermal) pions,
which can barely be separated experimentally from the pions from resonance
decay, show a very similar behaviour as do the radii extracted from kaon
interferometry. It is again a confirmation that kaon interferometry probes
the spatial and temporal characteristics of the fireball whereas pion
interferometry probes the halo due to resonances.

The general features of the momentum-dependent effective radii which were
presented in refs.\cite{bernd2,bernd3} for $S + S$ at 200 $GeV$ hold
also for $Pb + Pb$ at 160 $AGeV$. We stress four observations:

(i) In the case of pion interferometry, the resonance decays increase the
effective radii by almost the same factors as for $S + S$ at
200 $AGeV$ ($\sim 2.1$ in the longitudinal  and $\sim 1.4$ in
transverse direction).

(ii) The difference between the kaon and pion radii determines the size of
the resonance halo. Its maximum extension is 5.6 $fm$ in longitudinal
direction, 2.4 $fm$ in side- and 2.6 $fm$ in out directions (compare
corresponding values for $S+S$ at 200 $AGeV$ in table 1).   The halo size in
the longitudinal direction is larger because here we have stronger
retardation effects due to the delayed resonance decay because of
high fluid velocities.

(iii) We note that the Sinyukov-formula ({\it cf.} ref.\cite{loerstadt})
fails to describe our results although it was derived from hydrodynamical
assumptions. The reason for the difference is that
Sinyukov's formula was obtained for a longitudinally expanding source which is
better described in terms of the Bjorken scenario. In the case of $Pb+Pb$ we
have more stopping and thus a larger transverse expansion compared to
$S+S$. For $S + S$ at 200 $AGeV$ we found a maximum transverse velocity
$u_\perp^{max}(S)=0.43$ whereas for $Pb + Pb$ at 160 $AGeV$ we obtain
$u_\perp^{max}(Pb)=0.61$. Sinyukov's approach is therefore too rough,
and this results in the deviations. It also suggests that it is dangerous to
overstress the Bjorken approach because it can lead to misleading or
inconsistent
results.

(iv) The correlation functions are measurable with the experimental
resolution of the detector setups. Even the influence of the $\omega$ should
be visible.

In Fig. 3b we show our results for effective radii as a function of the
average transverse momentum $K_{\perp }$ of the pair for all pions at
$y_K=4.5 - y_{cm} =1.6$.  In particular, the effective longitudinal radii
$R_{\parallel }$ are evaluated in the longitudinal comoving system (LCMS) in
order to make a comparison to preliminary NA49 measurements \cite{QM95}
possible. Our calculations agree surprisingly well with the data.

To summarize: we have shown that preliminary data of the NA49 Collaboration
can be reproduced with a self-consistent three-dimensional relativistic
hydrodynamic description assuming an equation of state with a first
order phase transition. We have made predictions for rapidity spectra of
pions, kaons and protons for $Pb+Pb$ at 160 $AGeV$. Our data analysis
indicates a stronger stopping and an enhanced transverse flow in the case of
$Pb+Pb$ collisions. The high baryon density in the central region will have
a strong influence  on the chemistry of the fireball and can no longer be
neglected \cite{future}. Models dealing with weak probes and
particle abundancies need to take this into account.

Bose-Einstein correlation functions for pions and for kaons have also been
predicted. It is important to note that in the case of the pion
interferometry, the presence of a resonance halo increases the size of the
fireball in the central region by factors $\sim 2.1$ in longitudinal and
$\sim 1.4$ in transverse direction. The NA49 data on interferometry, which
were presented in terms of inverse radii, are surprisingly well described.

The results of this work constitute further evidence that heavy-ion collisions
in the SPS region show fluid dynamical behaviour and  that the
assumption of a first order phase transition from QGP to hadronic matter
is necessary in order to reproduce the data as was first pointed out in
\cite{udo}
and later confirmed by several independent groups \cite{many}.\\[3ex]

We would like to thank B.V. Jacak, J.P. Sullivan and N. Xu for many helpful
discussions. This work was supported by the Federal Minister of Research and
Technology under contract 06MR731, the Deutsche Forschungsgemeinschaft (DFG)
and the Gesellschaft f\"ur Schwerionenforschung (GSI).  B.R. Schlei
acknowledges a DFG postdoctoral fellowship and R.W. is indebted to
A. Capella for the hospitality extended to him at the LPTHE, Univ. Paris-Sud.

\newpage
\noindent
{\Large {\bf Figure Captions}}\\

\begin{description}
\item[Fig. 1a]  Rapidity spectra for negative hadrons ($h^{-}$), negative
pions ($\pi ^{-}$), negative thermal pions ($th.\:\pi ^{-}$), protons ($p$),
protons without those from $\Lambda $ decay ($p\backslash \Lambda
\rightarrow p$), negative kaons ($K^{-}$) and negative thermal kaons ($%
th.\:K^{-}$). The data points are preliminary results of the NA49
Collaboration for negative hadrons from central $Pb+Pb$ collisions at 160 $%
AGeV$. Circles (diamonds) stand for the measurements from the VTP2 (MTPC)
({\it cf.} refs.\cite{NA49_1},\cite{NA49_2}).

\item[Fig. 1b]  Transverse momentum spectra for negative hadrons ($h^{-}$),
negative pions ($\pi ^{-}$), negative thermal pions ($th.\:\pi ^{-}$),
protons ($p$), protons without those from $\Lambda $ decay ($p\backslash
\Lambda \rightarrow p$), negative kaons ($K^{-}$) and negative thermal kaons
($th.\:K^{-}$).

\item[Fig. 2]  Distortion of the Bose-Einstein correlation functions of
negatively charged pions through resonance decays. The contributions from
resonance decays are successively added to the correlation function of
thermal $\pi ^{-}$ (dotted lines ``1''). The resultant correlation functions
of all $\pi ^{-}$ are given by the solid lines.

\item[Fig. 3a]  Dependence of the longitudinal and transverse radii
(extensions of emission zones) extracted from Bose-Einstein correlation
functions on the rapidity $y_K$ and transverse average momentum $K_{\perp }$
of the pair. The curves are for all $\pi ^{-}$ (solid lines), thermally
produced $\pi ^{-}$ (dotted lines) and $K^{-}$ (dashed lines).

\item[Fig. 3b]  Effective radii extracted from Bose-Einstein correlation
functions as a function of the transverse average momentum $K_{\perp }$ of
the pair for all pions at rapidity $y_K=4.5 - y_{cm} =1.6$. The longitudinal
effective radii are shown in the LCMS. The data points are preliminary
results of the NA49 Collaboration \cite{QM95}.
\end{description}

\newpage
\vspace{1.5cm}

\noindent
{\Large {\bf Table Caption}}
\begin{description}
\item[Table 1]  Properties of initial fireball, extracted from a
hydrodynamical analysis of
the $S+S$ NA35 data \cite{wenig} and the $Pb+Pb$ NA49
data \cite{NA49_1},\cite{NA49_2}.
\end{description}

\newpage
\vspace{2.5cm}
\centerline{\huge \bf Table 1}
\label{tab1}

\begin{table}
\begin{center}
\begin{tabular}{|l|l|l|}
\hline
 & S+S & Pb+Pb \\
\hline
\multicolumn{3}{|c|}{\bf Fit parameters}\\
\hline
Thermal energy in the central fireball& 0.43 & 0.65 \\
\hline
Longitudinal extension of central fireball & 0.6 $fm$ & 1.2 $fm$ \\
\hline
Rapidity at edge of central fireball & 0.9 & 0.9 \\
\hline
Rapidity at maximum of initial baryon distribution & 0.82 & 0.60 \\
\hline
Width $\sigma$ of initial baryon $y$ distribution & 0.4 & 0.4 \\
\hline
\multicolumn{3}{|c|}{\bf Output}\\
\hline
fraction of baryons in central fireball& 0.49 & 0.73\\
\hline
initial energy density & 13 $GeV/fm^3$ & 20 $GeV/fm^3$ \\
\hline
max. lifetime & 6.9 $fm/c$ & 13.5 $fm/c$ \\
\hline
lifetime of QGP & 1.5 $fm/c$ & 3.4 $fm/c$ \\
\hline
max. halo-size in ``longitudinal'' direction & 2.5 $fm$ & 5.6 $fm$ \\
\hline
max. halo-size in ``side'' direction  & 1.0 $fm$ & 2.4 $fm$ \\
\hline
max. halo-size in ``out'' direction & 1.3 $fm$ & 2.6 $fm$ \\
\hline
\end{tabular}\\
\end{center}
\end{table}

\end{document}